\documentclass{scrartcl}
\usepackage{amsmath,amssymb,amsthm}
\usepackage{graphicx}
\usepackage{multirow}
\usepackage{authblk}

\newcommand{\rlbl}[1]{$^{[#1]}$}

\begin{document}

\title{Comparison of nanomechanical properties of in
vivo and in vitro keratin 8/18 networks\footnote{Preprint }}

\author[a]{Tobias Paust\thanks{Both authors contributed equally to this work}}
\author[a]{Anke Leitner$^\dag$}
\author[a]{Ulla Nolte}
\author[b]{Michael Beil}
\author[c]{Harald Herrmann}
\author[a]{Othmar Marti}
\affil[a]{\emph{Institute of Experimental Physics, Ulm University Ulm, 89069 Ulm, Germany}}
\affil[b]{\emph{Internal Medicine I, University Hospital, Albert-Einstein Allee 23, 89081 Ulm, Germany}}
\affil[c]{\emph{Division of Molecular Genetics, German Cancer Resarch Center, Heidelberg. Germany}}

\date{2009-10-01}
\maketitle

\begin{abstract}
  In our work we compare the mechanical properties of the extracted keratin cytoskeleton of pancreatic carcinoma cells with the mechanical properties of in vitro assembled keratin 8/18 networks. For this purpose we use microrheology measurements with embedded tracer beads. This method is a suitable tool, because the size of the beads compared to the meshsize of the network allows us to treat the network as a continuum. Observing the beads motion with a CCD-High-Speed-Camera then leads to the dynamic shear modulus. Our measurements show lower storage moduli with increasing distance between the rim of the nucleus and the bead, but no clear tendency for the loss modulus. The same measurement method applied to in vitro assembled keratin 8/18 networks shows different characteristics of storage and loss moduli. The storage modulus is one order of magnitude lower than that of the extracted cytoskeleton and the loss modulus is higher. We draw conclusions on the network topology of both keratin network types based on the mechanical behaviour.

\end{abstract}

\textbf{Keywords}: AFM, cytoskeleton, dynamic shear modulus, keratin 8/18, particle tracking microrheology


\section{Introduction}

The impressive active and passive mechanical performance of a cell acting as the smallest biological unit able to move, survive and replicate independently has been of much interest. Most importantly, the motility of malignant cells is an important prognosticator in cancer\rlbl{1}. The cells movement and response to external mechanical stimuli is a complex process mainly determined by its cytoskeleton. The cytoskeleton is a protein network extending from the cell membrane to the nucleus and filling large parts of the cytoplasm. It mostly consists of three different types of protein fibres: actin filaments, microtubuli and intermediate filaments\rlbl{4}. Actin filaments, with a persistence length of about 15 pm, are responsible for the movement of the cell, whereas the stiffer microtubuli with a persistence length of several millimetres\rlbl{5} are the transport pathways in the cell. Intermediate filaments are more flexible with a persistence length between 500 nm and 1000 nm\rlbl{12} (Keratin 8/18 $P_L=500~nm$, unpublished data), their function appears to be to provide the mechanical stiffness and stability of the cell. Thus they are
hypothesized to be the “stress-buffering-system”\rlbl{2} of cells.

Whereas actin filaments and microtubuli have been investigated in much detail\rlbl{4,5}, little is known about intermediate filaments. We have recently demonstrated that the intermediate filament network in pancreatic cancer cells, mostly consisting of keratins 8 and 18, is a crucial determinator for cancer cell motility\rlbl{13}. To further improve the understanding of the role of keratin networks in this disease we are now focusing on the nanomechanical properties of the keratin cytoskeleton. This paper presents two different approaches: In vivo measurements of the mechanical properties of the extracted cytoskeleton and measurements of in vitro assembled protein networks. Both approaches have different advantages that shall be discussed briefly.

As a top down approach it is useful to extract the keratin cytoskeleton from whole cells\rlbl{13}. Here the original architecture of the keratin network is conserved while other parts of the cell, like membrane, organels, actin, microtubuli etc. are removed. The advantage is obvious: the mechanical properties of the original keratin cytoskeleton can be investigated without any environmental influence while the local network topology is easily measured by SEM afterwards\rlbl{15}. But there is one major problem. The extraction does not remove all other proteins and some linker molecules, i.e. plectin, still remain in situ, especially in the crosslinking points of the network. For this purpose it is reasonable to have a look from the opposite point of view. The bottom up approach of in vitro assembled protein networks is not as unnatural as it may sound. Especially the cytoskeleton is a part of the cell with several components fulfilling all their special tasks independently\rlbl{1}. Hence it is quite useful to analyse the mechanical properties of parts of the cytoskeleton as isolated functional modules. Additionally it is possible to create well defined networks in vitro without any undesirable components. Determining the viscoelastic and mechanical properties of these in vitro assembled networks should show comparable results to extracted cell networks.

Microrheology is a suitable tool for measurements of the mechanical properties of both the extracted cytoskeleton and of in vitro assembled keratin networks. The possibility to measure over an extended range only by observing thermal fluctuation is an advantage over the traditional measurements where mechanical stress is applied and the frequency range is limited\rlbl{9}. In our model micron-sized spherical particles, act as probes, which scan their environment driven by the thermal forces (Brownian motion) dependent on the viscoelastic behaviour of the surrounding network. By means of particle tracking the Brownian motion of several beads can be detected over a wide range of frequencies (up to 5000 Hz with a high-speed CCD-Camera) with a high accuracy in resolution (down to 5 nm). This allows to calculate the complex shear moduli which give an insight to the properties of the network and enables to show the role of the intermediate filaments in the cell. In that way general statement about network dynamics are possible\rlbl{7,9,19}.

\section{Results and Discussion}

\subsection{Shear moduli of intermediate filament networks}

Typical AFM\rlbl{20} pictures of the in vitro assembled network and the extracted cytoskeleton are shown in Figure 1. One can clearly see structural differences between the two kind of networks.

In this work the shear modulus of the extracted keratin cytoskeleton (K8/18) from Panc 1 cells is compared to the shear modulus of in vitro assembled keratin 8/18 networks. Both data sets from microrheology measurements are evaluated with the model relating stochastic movement to mechanical frequency dependent moduli.

\subsection{Theoretical Aspects}

The derivation of the complex shear modulus from the average motion of beads was first described by Mason in the late 90's\rlbl{9}. The model is based on the generalized Stokes-Einstein equation and provides a good way to determine the linear viscoelasticity of complex fluids. To derive the moduli we assume that the filament network is treated as a viscoelastic, incompressible, isotropic fluid. It therefore is assumed to be a continuum and the beads' inertia was neglected, too. Around the beads' surface there are noslip boundary conditions which are a good approximation in the frequency range covered by our experiment. A brief sketch of the derivation of the shear moduli is given below\rlbl{7,8,9,10}.

The motion of a single bead with radius $r$ in a viscoelastic medium can be described by a generalized Langevin equation\rlbl{11}:

\begin{equation*}
  m \dot{v}(t) = f_R(t)-\int_0^t dt' \zeta(t-t')v(t'),
\end{equation*}

with $m$ and $\dot{v}$ being the mass of the bead and its acceleration. The memory function $\zeta$ describes the response of the incompressible complex fluid. The function $f_R(t)$ contains the stochastic forces of the viscous fluid. Causality and the usage of the equipartition theorem of thermal energy relates the local memory function to the velocity of the bead. The complex viscosity $\eta(\omega)$ can be related to the memory function $\zeta$ by using the Stokes relation:

\begin{equation*}
  \eta(\omega) = \frac{\zeta(\omega)}{6\pi r}
\end{equation*}

With $G(\omega)=i \omega \eta(\omega)$ the complex shear modulus can be calculated in terms of a unilateral Fourier transform

\begin{equation*}
  G(\omega) = \frac{k_B T}{\pi\, r\, i\, \omega F_u\left\{\left<\Delta r^2(t)\right>\right\}}
\end{equation*}

This equation represents a generalization of the Stokes-Einstein equation in the Fourier domain. The evaluation of the Fourier term can be done by an estimation of the transforms by expanding $\left<r^2(1/\omega)\right>$, the mean square displacement, locally around the frequency of interest using the power law expansion. This evaluation leads to a relation which is suitable for analytic computation. Hence, we find

\begin{equation*}
  \left| G(\omega)\right| = \frac{k_B T}{\pi\, r \left< \Delta r^2\left(1/\omega\right)\right>\, \Gamma\left[ 1+\alpha(\omega)\right]}
\end{equation*}

with $\alpha$ the power law exponent which depends on the logarithmic
slope of the mean square displacement (MSD) $\left<\Delta r^2\left(1/\omega\right)\right>$. Since $G(\omega)$ in general is a complex function, it can be split into the real and the imaginary part. They are related by the Kramers-Kronig-Relation. $G'(\omega)$, the so called storage modulus, is the real part and describes the dissipation-free spring-like behaviour. $G''(\omega)$, the loss modulus, is the imaginary part and gives information about the liquid-like dissipative behaviour of the material. For the shear moduli we achieve

\begin{align*}
  G'(\omega) &= \left| G(\omega)\right| \cdot \cos\left(\pi\, \alpha(\omega)/2\right)\\
  G''(\omega) &= \left| G(\omega)\right| \cdot \sin\left(\pi\, \alpha(\omega)/2\right)
\end{align*}

In simple viscous fluids the observed MSD becomes proportional to $\tau$ and the complex modulus purely imaginary with $G(\omega) = i \omega \eta$ proportional to the macroscopic shear viscosity. For a simple elastic solid the MSD is lag-time independent. A viscoelastic material shows an intermediate form, where the storage modulus dominates at low frequencies and the loss modulus at high ones.

\subsection{Comparison of shear moduli}

The human Keratin 8/18 network was analysed in vitro to compare the mechanical properties with those from the extracted keratin cytoskleton of a cell. Figure 2 shows storage and loss moduli for a fixed frequency $\omega=10^{3/2}~Hz$. One can clearly see the dependency between the distance rim of the nucleus-bead and the moduli. The farther away the bead is from the nucleus rim the weaker is the surrounding network. Compared to the extracted cytoskeleton the storage modulus of the in vitro assembled network is about one order of magnitude lower whereas the loss modulus of the in vitro assembled network is higher than that of the extracted keratin cytoskeleton. Possible reasons are mainly the differences in the construction of the two networks and the differences in the bead interaction.

On the one hand the extracted cytoskeleton is a dense crosslinked network with chemical bondings at the branching points as displayed in figure 1 on the right side. On the other hand the in vitro polymerised network is entangled with at most few intersections (figure 1 left side). The force connecting filaments to an entanglement originates from friction forces and not from chemical bonding and leads therefore to a weaker connection point. This difference can be clearly seen in figure 3. Two point clouds from typical measurements are displayed as iso surface representations of the 3d position distribution. Figure 3a shows the density cloud of a bead embedded in the extracted cytoskeleton. The bead is fluctuating almost homogeneously in all directions. By contrast figure 3b shows the iso surface representation of the 3d position distribution of a bead in the in vitro assembled network. The bead is fluctuating more freely. It can circle a filament (small picture) or jump from one mesh to another (supporting data). From this structural difference the characteristics of the moduli can be explained with a possible reptation-like behaviour in the entangled network\rlbl{16}. This leads to a higher loss modulus and a lower storage modulus because of the dissipation in the network. By contrast in the crosslinked network the storage modulus is higher and the dissipation is lower because reptation like movements are suppressed.

The different behaviour might be due to differences in the assembly process in cells and in vitro. In the cell a directed assembly process takes place whereas the in vitro polymerisation is a self assembly process starting everywhere in the volume. The network in the cell is continuously constructed with the help of many players like helper proteins or linker proteins\rlbl{18}. Additionally there may be a preferred direction of assembly because of cell movement or a prestress from the membrane. All these factors indicate a different topology compared to the in vitro polymerised network. Here the assembly is a stochastic temperature driven process. There is no influence from the surrounding environment or a directed development of the network. Just two filament ends meeting coincidentally lead to a junction and two crossing filaments can lead to an entanglement.

The interaction of the bead with the network is strongly dependent on the polymerisation process. During the self assembly of the in vitro polymerised network the bead is continuously fluctuating. This movement leads to the formation of a cavity around the bead where no polymerisation takes place. Thus the bead is not as strongly embedded in the network as in case of the cytoskeleton. There the bead is integrated in the network by strong connections of single filaments with the bead. Another aspect is the keratin concentration of 1 mg/ml (see Experimental section) in the network. This leads to an estimated mesh size of 400 nm in the in vitro polymerised network and is comparable to the mesh size of the extracted  cytoskeleton\rlbl{15}.

\section{Conclusion}

Our experiments compare the dynamic shear moduli of the extracted keratin cytoskleton with those of in vitro assembled keratin networks. It was shown that the dynamic shear modulus of the keratin cytoskeleton is decreasing significantly with increasing distance to the nucleus. In comparison the storage modulus of the in vitro assembled network is about one order of magnitude lower whereas the loss modulus of the in vitro assembled network is higher than that of the extracted keratin cytoskeleton. One possible explanation is based on the different topology of the two types of networks. The in vitro assembled networks is dominated by entanglements. The movement of the filaments is probably reptation-like causing dissipation in the network. This leads to a higher loss modulus and a lower storage modulus. The extracted keratin cytoskeleton is a crosslinked network, where reptation-like movements are suppressed. That leads to an increasing storage modulus and, because of the lower dissipation, to a lower loss modulus.

Additionally the presented work shows that rheology measurements from 0.1~Hz up to 5~kHz are possible and $G'$ and $G''$ can be determined in a range from $10^{−5}~Pa$ to $10~Pa$. Here microrheology is a suitable tool for investigation of the link between structure and function.

The comparison of measurements of in vitro assembled networks with the extracted keratin cytoskeleton of living cells shows that the bottom up approach is a promising way for further investigations. Indeed it is very important to take the structural differences into account.

\section{Experimental Section}

Human Keratin 8 and Keratin 18 were expressed and purified as described in Harald Herrmann et al 1999\rlbl{17}. The assembly protocol was as follows\rlbl{3}: Keratin 8 and Keratin 18 were mixed in 8 M Urea in 1:1 ratio. Afterwards the urea concentration was lowered stepwise by dialysis first into buffer1 (4 mM urea, 10 mM Tris, 2 mM DTT, pH 8.5) then buffer2 (2mM urea, 10 mM Tris, 2 mM DTT, pH 8.5) and finally buffer3 (0 mM urea, 10 mM Tris, 2 mM DTT, pH 8.5) at room temperature. Dialysis was continued overnight into dialysis buffer (2 mM Tris, 1 mM DTT, pH 9.0) at $4^\circ$C. For starting the assembly an equal volume of 2 x assemblypuffer (20 mM Tris, pH 7.0) was added. The assembly concentration was 1 mg/ml.

For the sample preparation of the AFM measurements of the in vitro assembled network the assembly was diluted to a concentration of 0.005 mg/ml. $50~\mu$l of the keratin assembly was applied to a freshly cleaved mica surface, fixed with a fixation buffer (0.2 \% Glutaralduhyde) and carefully washed with filament buffer (equal volumes of dialysis buffer and 2 x assemblybuffer). The extracted cytoskeleton was fixed with 4\% glutaraldehyde and gradually dehydrated and subjected to critically point drying. AFM measurements were done with a scan frequency of 0.5 Hz using cantilevers for tapping mode from olympus (OMCL-AC240TS-W2) with a nominal spring constant of 2 N/m. The pictures were recorded using an MFP-3D from asylum research and processed with WSxM 4.0 Develop 12.21\rlbl{14}.

The protein concentration of the assembly for the microrheology measurements was 1 mg/ml. Additionally 1$~\mu$m spheres were added in a concentration of 1 \% by weight in millipore water to $100~\mu$l protein solution before starting the assembly.

For the measurement of the brownian motion beads of a diameter of 500 nm and $1~\mu$m were used in an extracted keratin network. Living carcinoma cells incorporate the beads and embed them into their cytoskeleton. After the extraction of the cell remains the keratin network, the nucleus and the incorporated beads. For further details on cell extraction we refer to Beil et al. 2003\rlbl{13}. A microscope setup with additional high-speed camera was used to track the embedded beads with a frequency of 5000 Hz at a time window of 3.2 sec and a resolution of 256x256 pixels. The usage of the camera allows the tracking of several beads simultaneously in three dimensions (manuscript in preparation) and therefore has the advantage of the measurement of beads at the same conditions.

\section{Acknowledgements}

We thank the DFG SFB518 (TP) and Project MA1297/10-1 (AL) for financial support, Andreas H\"{a}u{\ss}ler, Carlo Di Giambattista and Sarah Pomiersky for discussion and measurements and all other members of the Institute of Experimental Physics, Ulm Univerity, Paul Walther and the ZE Elektronenmikroskopie, Ulm University, and the members of the Division of Molecular Genetics from the DKFZ Heidelberg.

Special thanks to Prof. Adler and the Department of Internal Medicine I, University Hospital Ulm, and to Prof. Mizaikoff and the Institute of Analytical and Bioanalytical Chemistry, Ulm University, for sharing laboratories during constructions.

\newpage

\textbf{References}

\

\begin{enumerate}
  \renewcommand{\labelenumi}{[{\arabic{enumi}}]}
\item A. R. Bausch, K. Kroy, Nature physics 2006, 2, 231-238;
\item  H. Herrmann, U. Aebi, Annu. Rev. Biochem.2004, 73, 749-789;
\item H. Herrmann, T. Wedig, R. M. Porter, E. B. Lane, U. Aebi, J. Struct. Biol.
2002, 137, 82-96;
\item E. Fuchs, K. Weber, Annu. Rev. Biochem. 1994, 63, 345-382;
\item T. Yanagida, M. Nakase, K. Nishiyama, F. Oosawa, Nature 1984, 301,
58;
\item F. Gittes, B. Mickey, J. Nettleton, J. Howard, J. Cell Biol. 1993, 120, 923;
\item T. G. Mason, Rheol Acta 2000, 39, 371-378;
\item F. Gittes, B. Schnurr, P.D. Olmsted, F.C. MacKintosh, C. F. Schmidt,
Phys. Rev. Letter. 1997, 79, 3286-3289;
\item T. G. Mason, D. A. Weitz, Phys. Rev. Letter. 1995, 74, 1250-1253;
\item R. B. Bird, R. C. Armstrong, O. Hassager, Dynamics of polymer liquids
1977, Wiley, New York;
\item J.-P Hansen, I. R. McDonald., Theory of simple liquids 1986, Academic
Press, London;
\item N. M\"{u}cke, L. Kreplak, R. Kirmse, T Wedig H. Herrmann, U. Aebi, J.
Langowski, J. Mol. Biol. 2004, 335, 1241-1250;
\item M. Beil, A. Micoulet, G. v. Wichert, S. Paschke, P. Walther, M. Bishr
Omary, P.P. Van Veldhoven, U. Gern, E. Wolff-Hieber, J. Eggermann, J.
Waltenberger, G. Adler, J. Spatz, T. Seufferlein, Nat. Cell Biol. 2003, 5,
803-811;
\item I. Horcas, R. Fernandez, J. M. Gomez-Rodriguez, J. Colchero, J.
Gomez-Herrero, A. M. Baro, Rev. Sci. Instrum. 2007, 78, 013705;
\item M. Beil, H. Braxmeier, F. Fleischer, V. Schmidt, P. Walther, Journal of
Microscopy 2005, 220, 84-95;
\item P. G. De Gennes, J. Chem. Phys. 1971, 55, 572-579;
\item H. Herrmann, M. H\"{a}ner, M. Brettel, N. Ku, U. Aebi, J. Mol. Biol. 1999,
286, 1403-1420;
\item R. Windoffer, S. W\"{o}ll, P. Strnad, R. E. Leube, Mol. Biol. Cell. 2004, 15,
2436-2448;
\item J. C. Crocker, B. D. Hoffman, Meth. Cell Biol. 2007, 83, 141-178;
\item For an overview see B. Bhushan, O. Marti, Handbook of
nanotechnology, 2007, Springer, New York, and references there in;

\end{enumerate}

\newpage

\section*{Figure Captions}

\begin{enumerate}
\renewcommand{\labelenumi}{Figure \arabic{enumi}.}

\item  a) AFM picture of the in vitro assembled keratin network, 7.5~$\mu$m~x~7.5~$\mu$m topography b) AFM picture of the extracted cytoskeleton, $2~\mu$m~x~2~$\mu$m topography.

\item $G'$, $G''$ as function of distance to the rim of the nucleus for a fixed frequency $\omega=10^{1.5}$~Hz. $G'$ (empty squares) and $G''$ (empty circles) of the extracted cytoskeleton and $G'$ (full squares) and $G''$ (full circles) of the in vitro assembled network. In this sketch the shear moduli of the in vitro assembled network are constant, because of the lack of a nucleus.

\item a) Iso surface representation of 3d position distribution of a bead in the in vitro assembled network. b) Iso surface representation of 3d position distribution of a bead in the extracted cytoskeleton. Iso surface representation of 3d position distribution shows all point with an equal probability to find the particle.

\end{enumerate}

\newpage

\includegraphics[width=\textwidth,height=0.8\textheight,keepaspectratio]{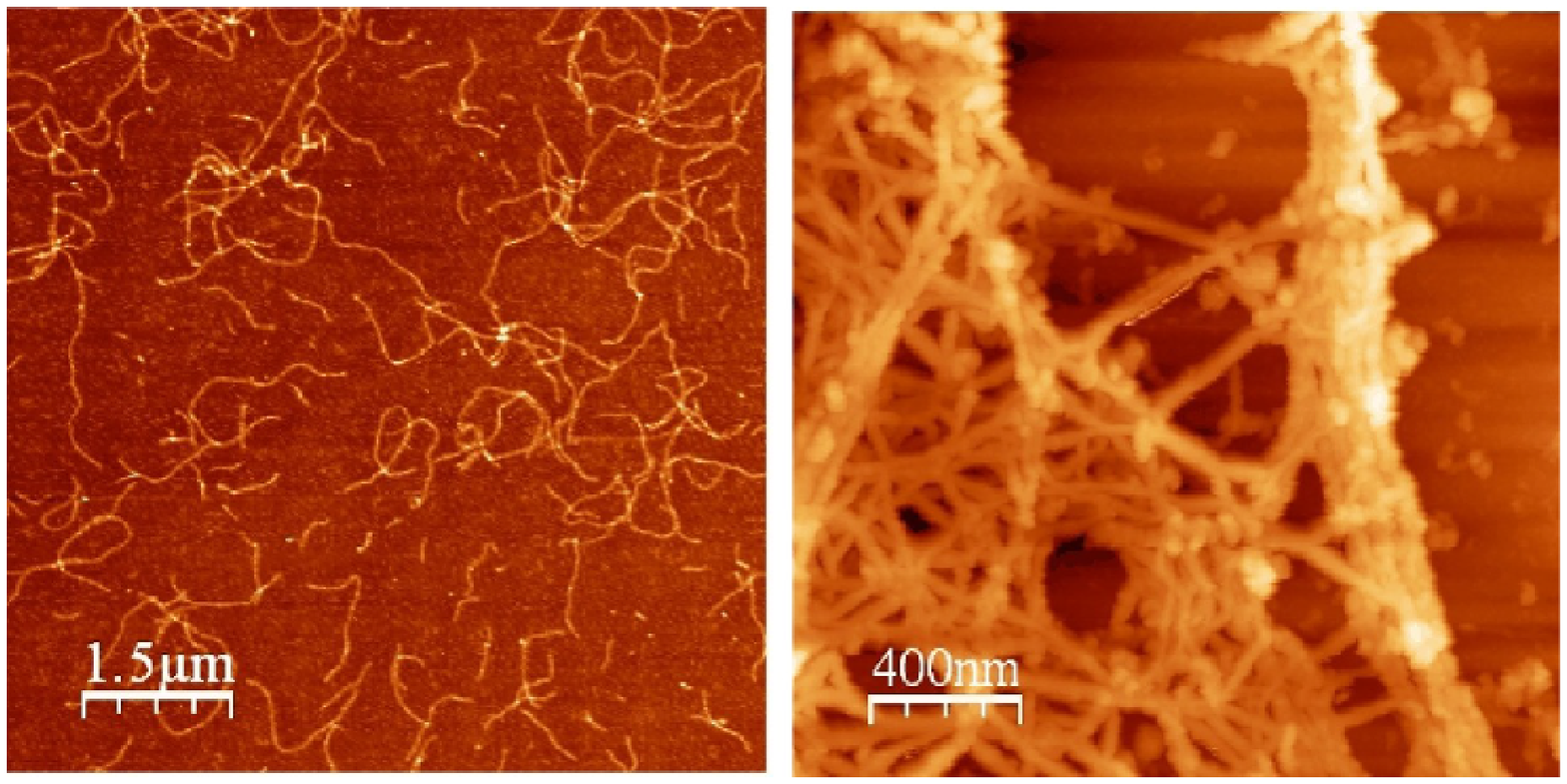}
\vfill
\Large{Figure 1, Comparison of nanomechanical properties of in
vivo and in vitro keratin 8/18 networks Tobias Paust, Anke Leitner, Ulla Nolte, Michael Beil, Harald Herrmann, Othmar Marti}

\newpage

\includegraphics[width=\textwidth,height=0.8\textheight,keepaspectratio]{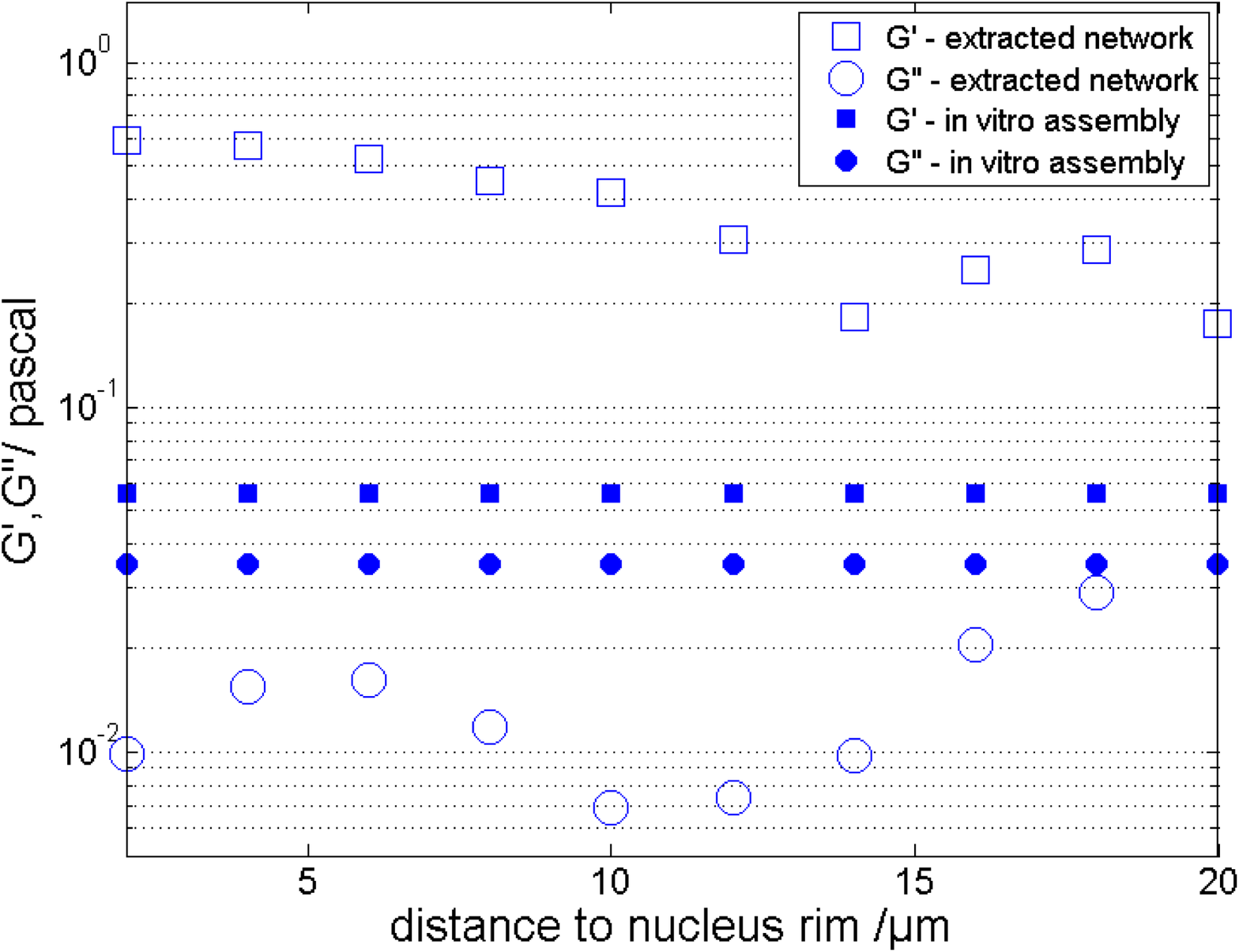}
\vfill
\Large{Figure 2, Comparison of nanomechanical properties of in
vivo and in vitro keratin 8/18 networks Tobias Paust, Anke Leitner, Ulla Nolte, Michael Beil, Harald Herrmann, Othmar Marti}

\newpage

\includegraphics[width=\textwidth,height=0.8\textheight,keepaspectratio]{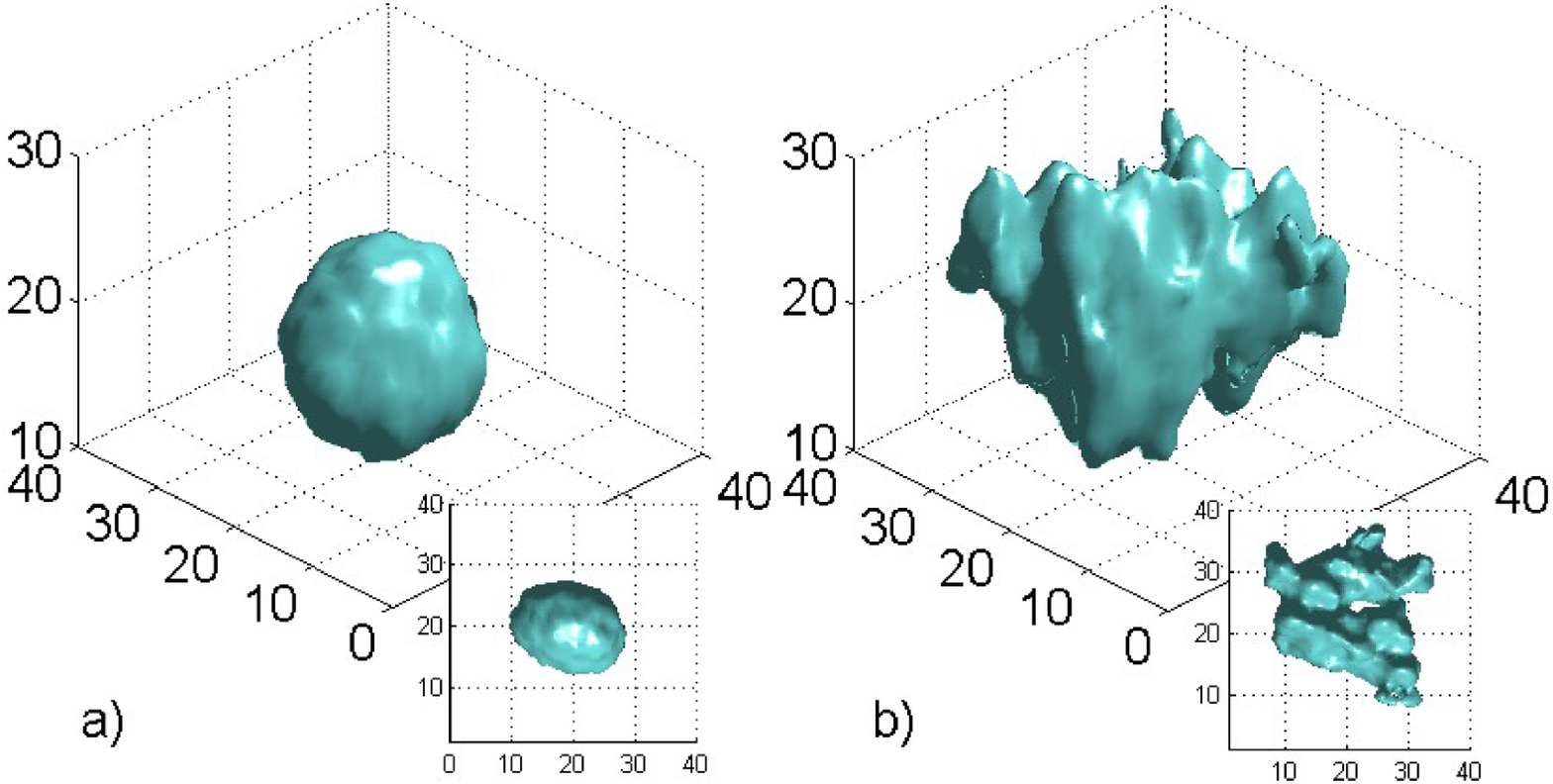}
\vfill
\Large{Figure 3, Comparison of nanomechanical properties of in
vivo and in vitro keratin 8/18 networks Tobias Paust, Anke Leitner, Ulla Nolte, Michael Beil, Harald Herrmann, Othmar Marti}

\newpage
\section*{Supplemental Data:}

\includegraphics[width=\textwidth,height=0.8\textheight,keepaspectratio]{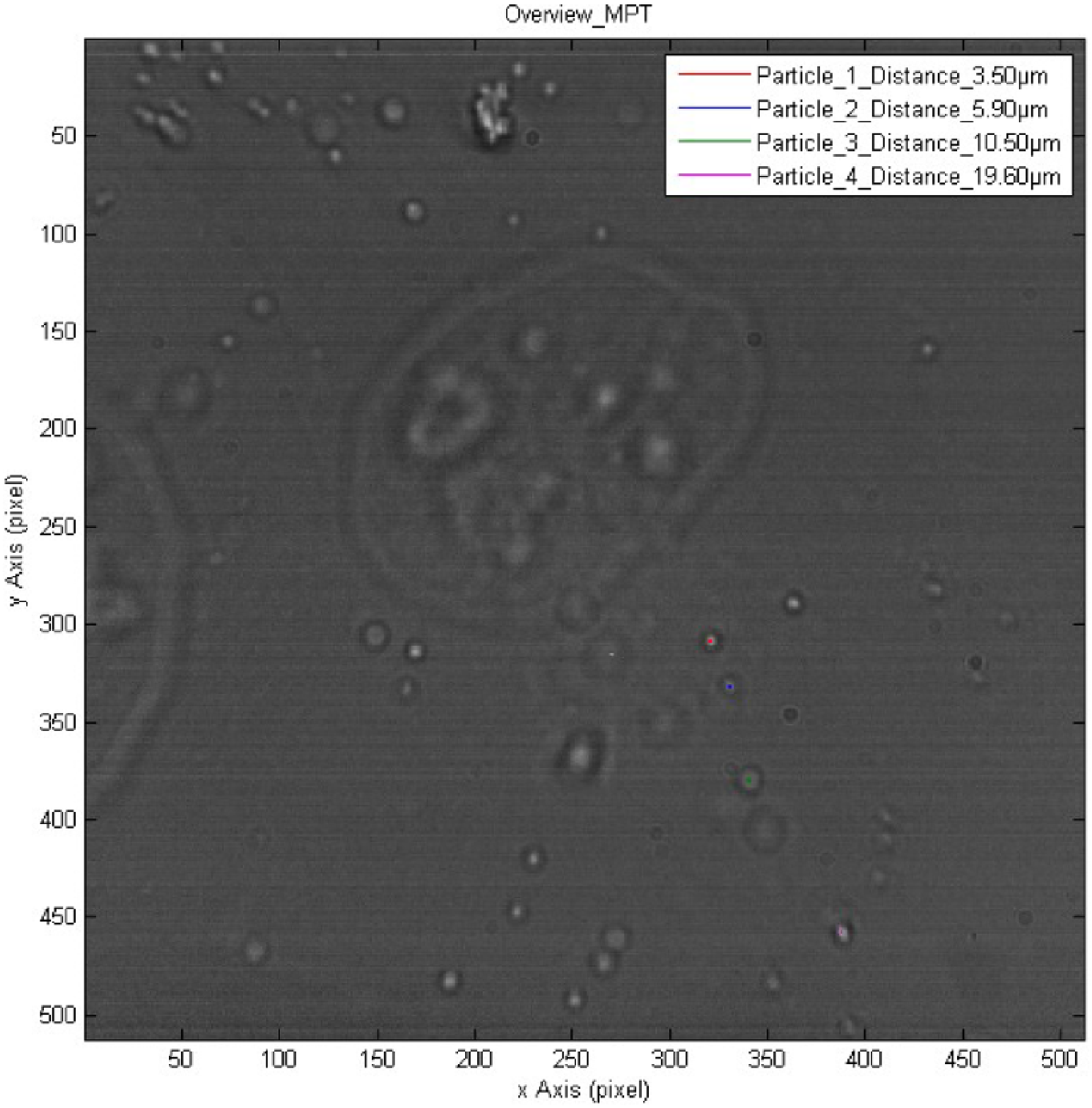}
\vfill

\Large{Extracted network with four beads of different distance to the rim of nucleus}

\large{Supplemental data, Comparison of nanomechanical properties of in
vivo and in vitro keratin 8/18 networks Tobias Paust, Anke Leitner, Ulla Nolte, Michael Beil, Harald Herrmann, Othmar Marti}

\newpage

\includegraphics[width=\textwidth,height=0.8\textheight,keepaspectratio]{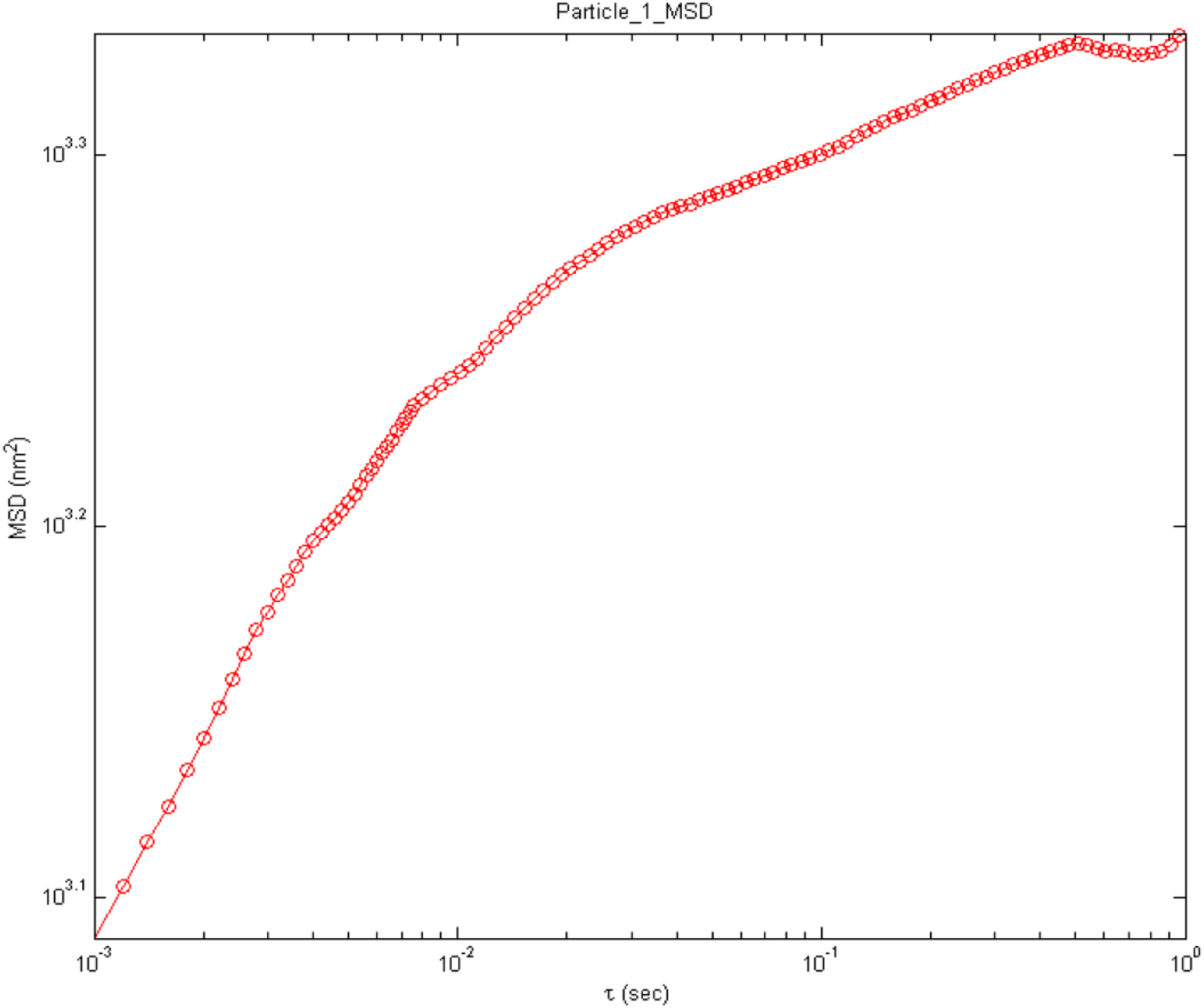}
\vfill

\Large{Exemplary Measurement: Distance to the rim of the nucleus 3.5~$\mu$m:}

\Large{Mean-square displacement}

\large{Supplemental data, Comparison of nanomechanical properties of in
vivo and in vitro keratin 8/18 networks Tobias Paust, Anke Leitner, Ulla Nolte, Michael Beil, Harald Herrmann, Othmar Marti}

\newpage

\includegraphics[width=\textwidth,height=0.8\textheight,keepaspectratio]{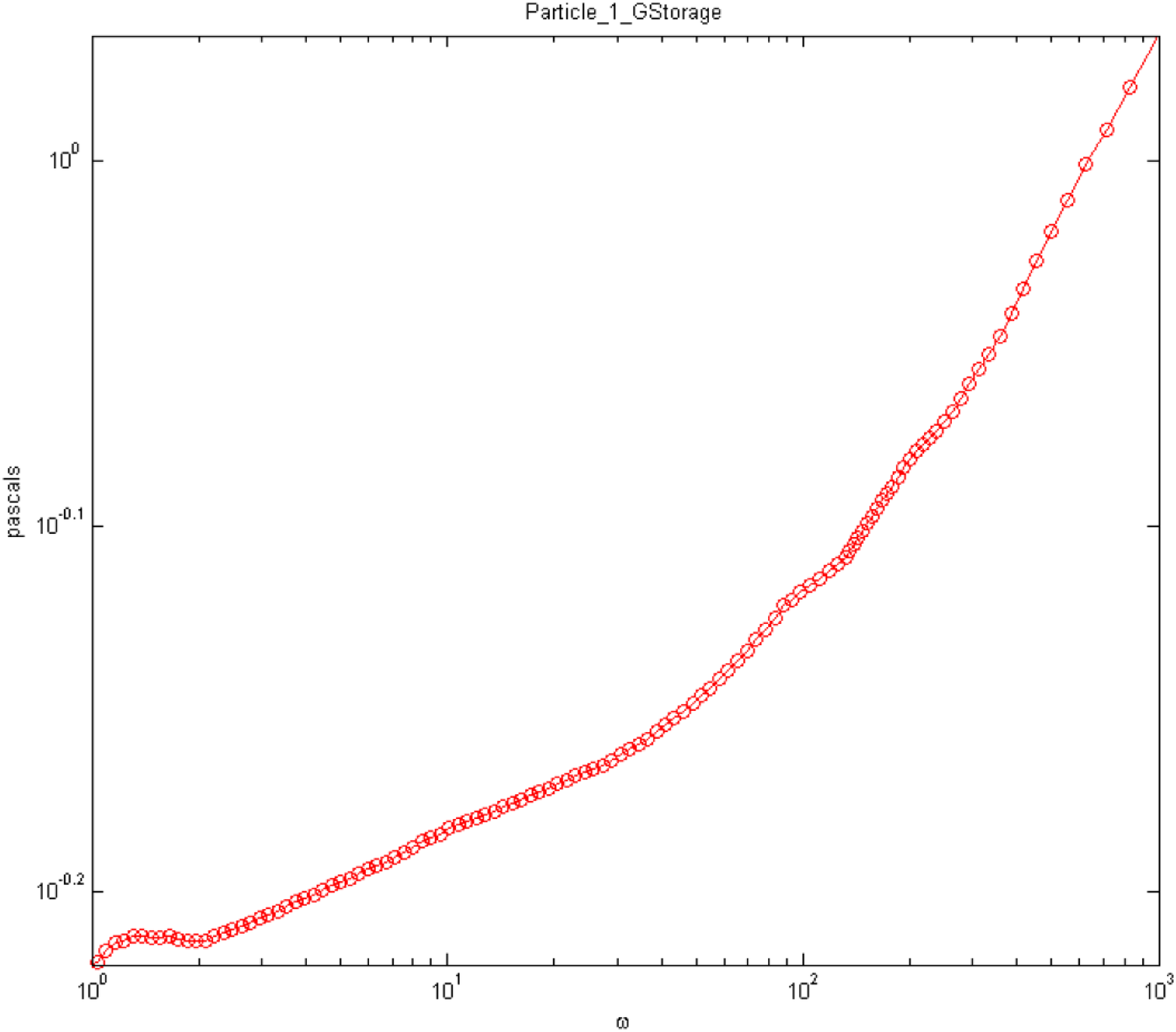}
\vfill

\Large{Exemplary Measurement: Distance to the rim of the nucleus 3.5~$\mu$m:}

\Large{Storage modulus}

\large{Supplemental data, Comparison of nanomechanical properties of in
vivo and in vitro keratin 8/18 networks Tobias Paust, Anke Leitner, Ulla Nolte, Michael Beil, Harald Herrmann, Othmar Marti}

\newpage

\includegraphics[width=\textwidth,height=0.8\textheight,keepaspectratio]{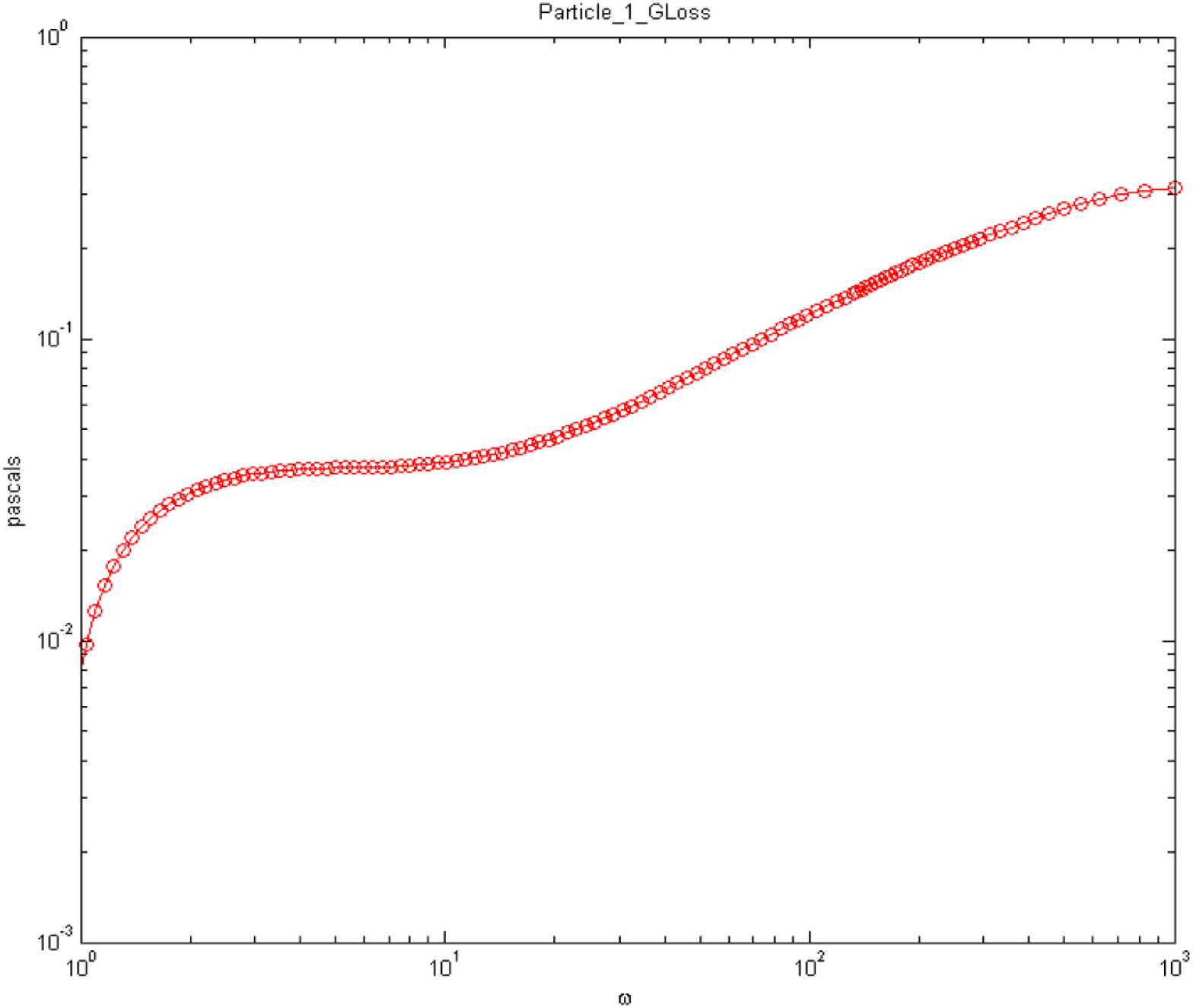}
\vfill

\Large{Exemplary Measurement: Distance to the rim of the nucleus 3.5~$\mu$m:}

\Large{Loss modulus}

\large{Supplemental data, Comparison of nanomechanical properties of in
vivo and in vitro keratin 8/18 networks Tobias Paust, Anke Leitner, Ulla Nolte, Michael Beil, Harald Herrmann, Othmar Marti}

\newpage

\includegraphics[width=\textwidth,height=0.8\textheight,keepaspectratio]{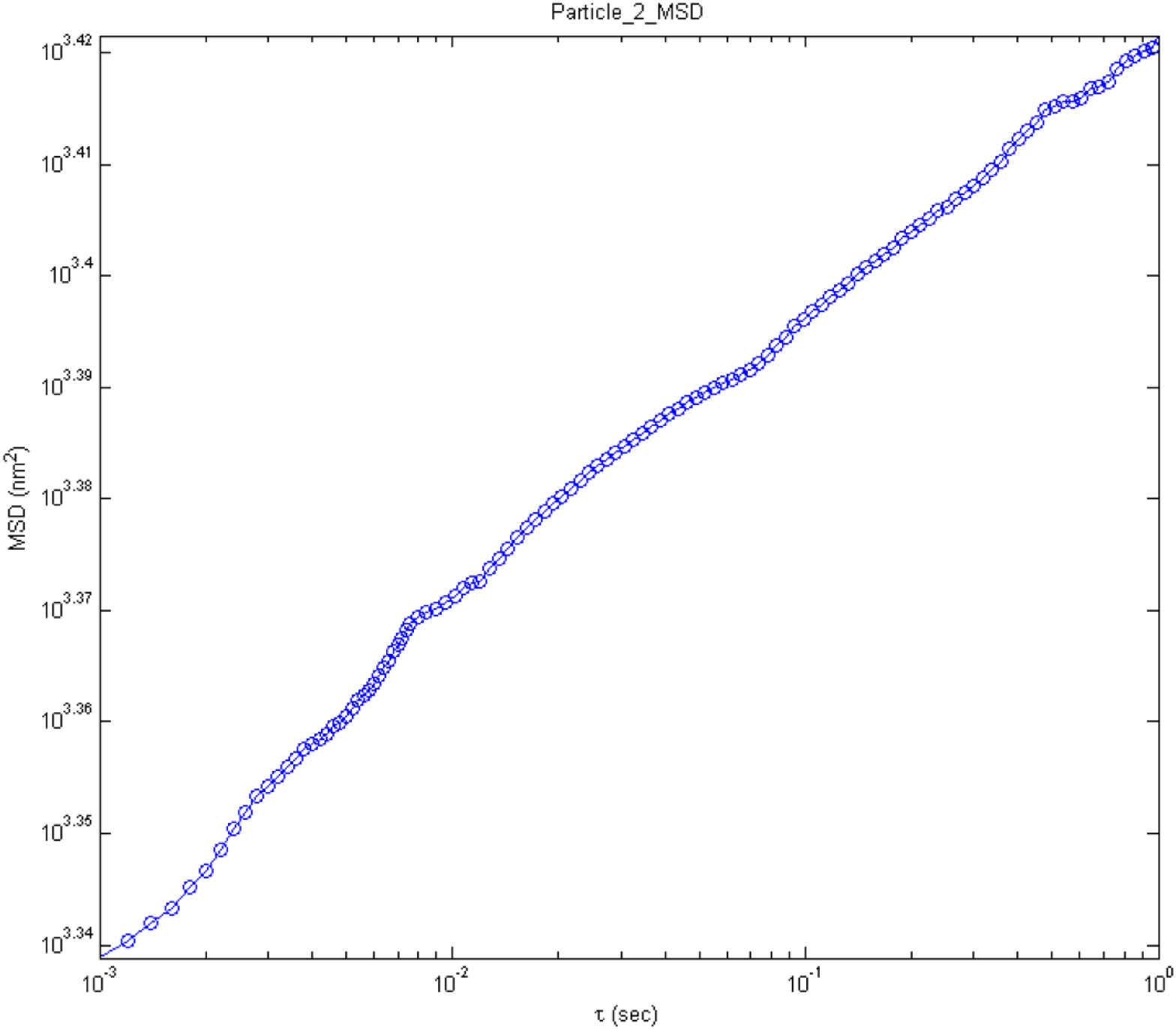}
\vfill

\Large{Exemplary Measurement: Distance to the rim of the nucleus 5.9~$\mu$m:}

\Large{Mean-square displacement}

\large{Supplemental data, Comparison of nanomechanical properties of in
vivo and in vitro keratin 8/18 networks Tobias Paust, Anke Leitner, Ulla Nolte, Michael Beil, Harald Herrmann, Othmar Marti}

\newpage

\includegraphics[width=\textwidth,height=0.8\textheight,keepaspectratio]{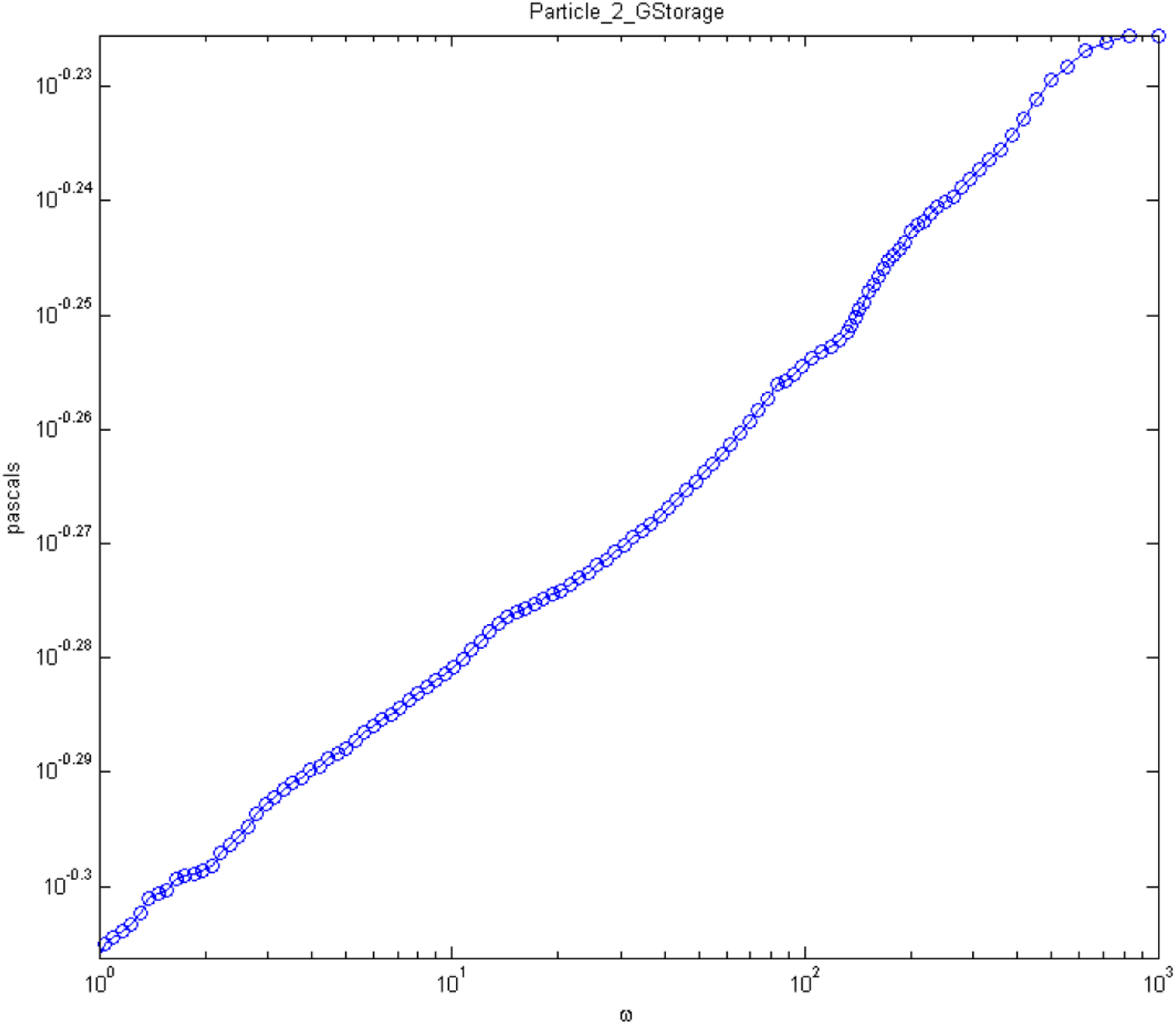}
\vfill

\Large{Exemplary Measurement: Distance to the rim of the nucleus 5.9~$\mu$m:}

\Large{Storage modulus}

\large{Supplemental data, Comparison of nanomechanical properties of in
vivo and in vitro keratin 8/18 networks Tobias Paust, Anke Leitner, Ulla Nolte, Michael Beil, Harald Herrmann, Othmar Marti}

\newpage

\includegraphics[width=\textwidth,height=0.8\textheight,keepaspectratio]{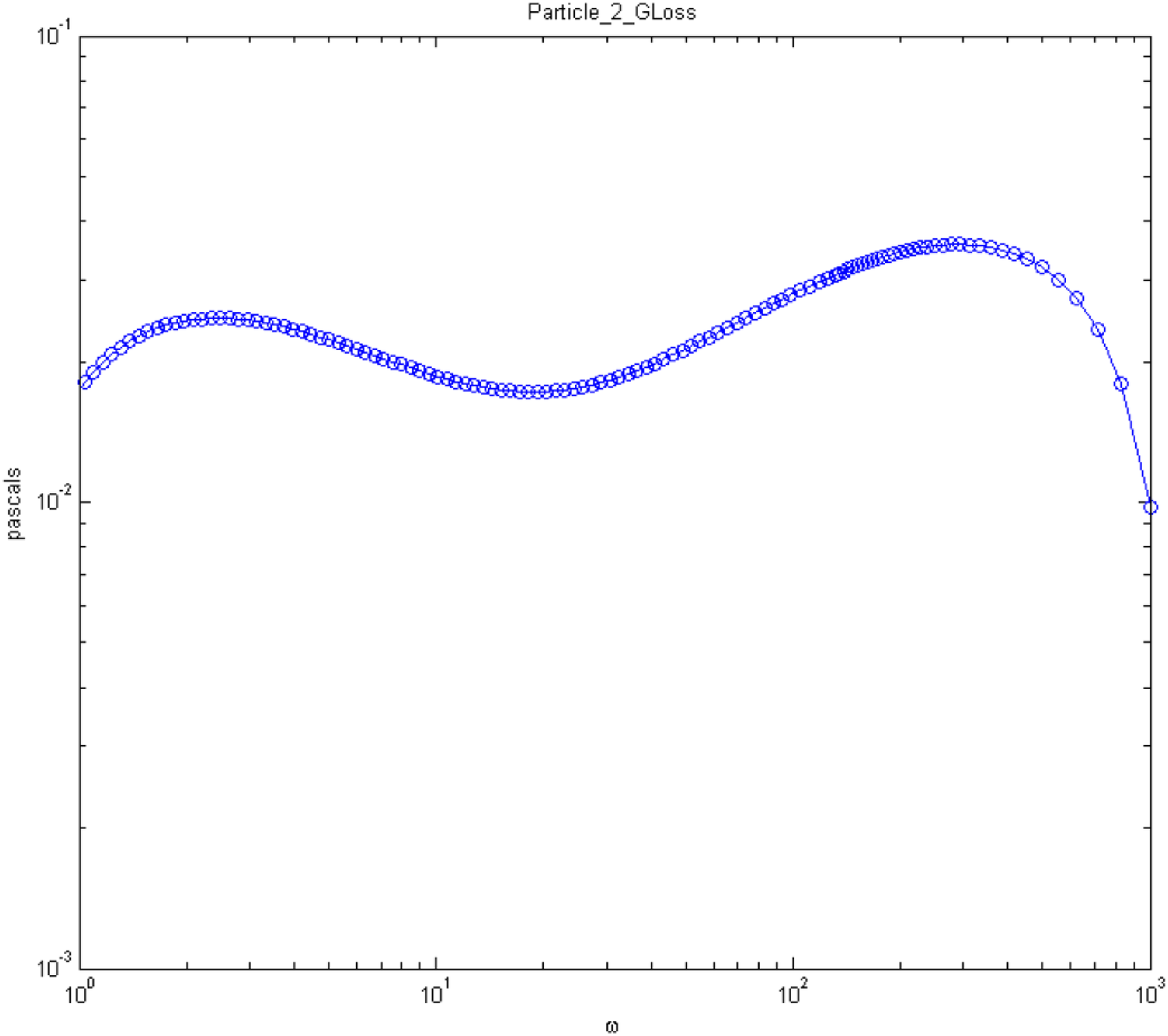}
\vfill

\Large{Exemplary Measurement: Distance to the rim of the nucleus 5.9~$\mu$m:}

\Large{Loss modulus}

\large{Supplemental data, Comparison of nanomechanical properties of in
vivo and in vitro keratin 8/18 networks Tobias Paust, Anke Leitner, Ulla Nolte, Michael Beil, Harald Herrmann, Othmar Marti}

\newpage

\includegraphics[width=\textwidth,height=0.8\textheight,keepaspectratio]{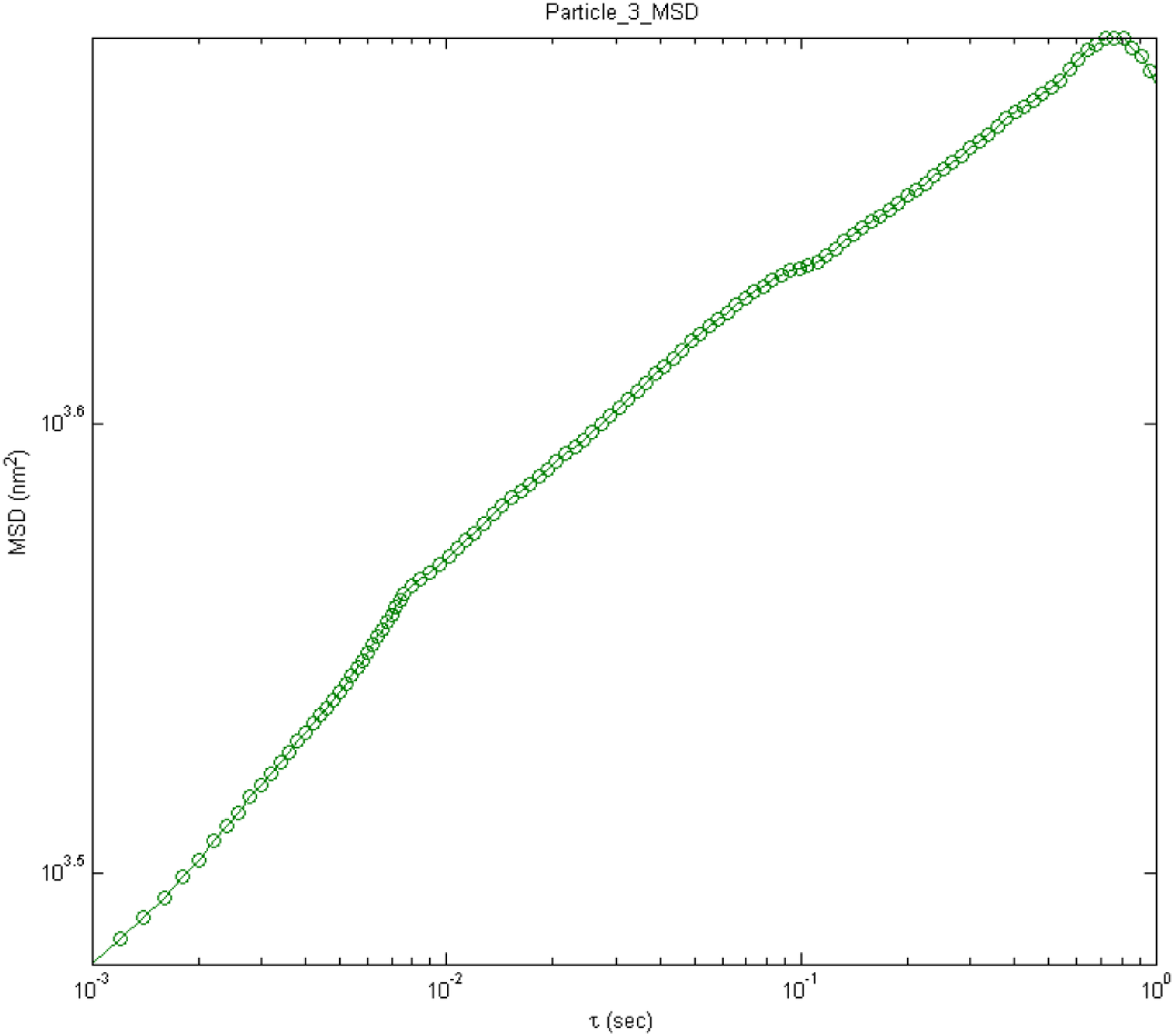}
\vfill

\Large{Exemplary Measurement: Distance to the rim of the nucleus 10.5~$\mu$m:}

\Large{Mean-square displacement}

\large{Supplemental data, Comparison of nanomechanical properties of in
vivo and in vitro keratin 8/18 networks Tobias Paust, Anke Leitner, Ulla Nolte, Michael Beil, Harald Herrmann, Othmar Marti}

\newpage

\includegraphics[width=\textwidth,height=0.8\textheight,keepaspectratio]{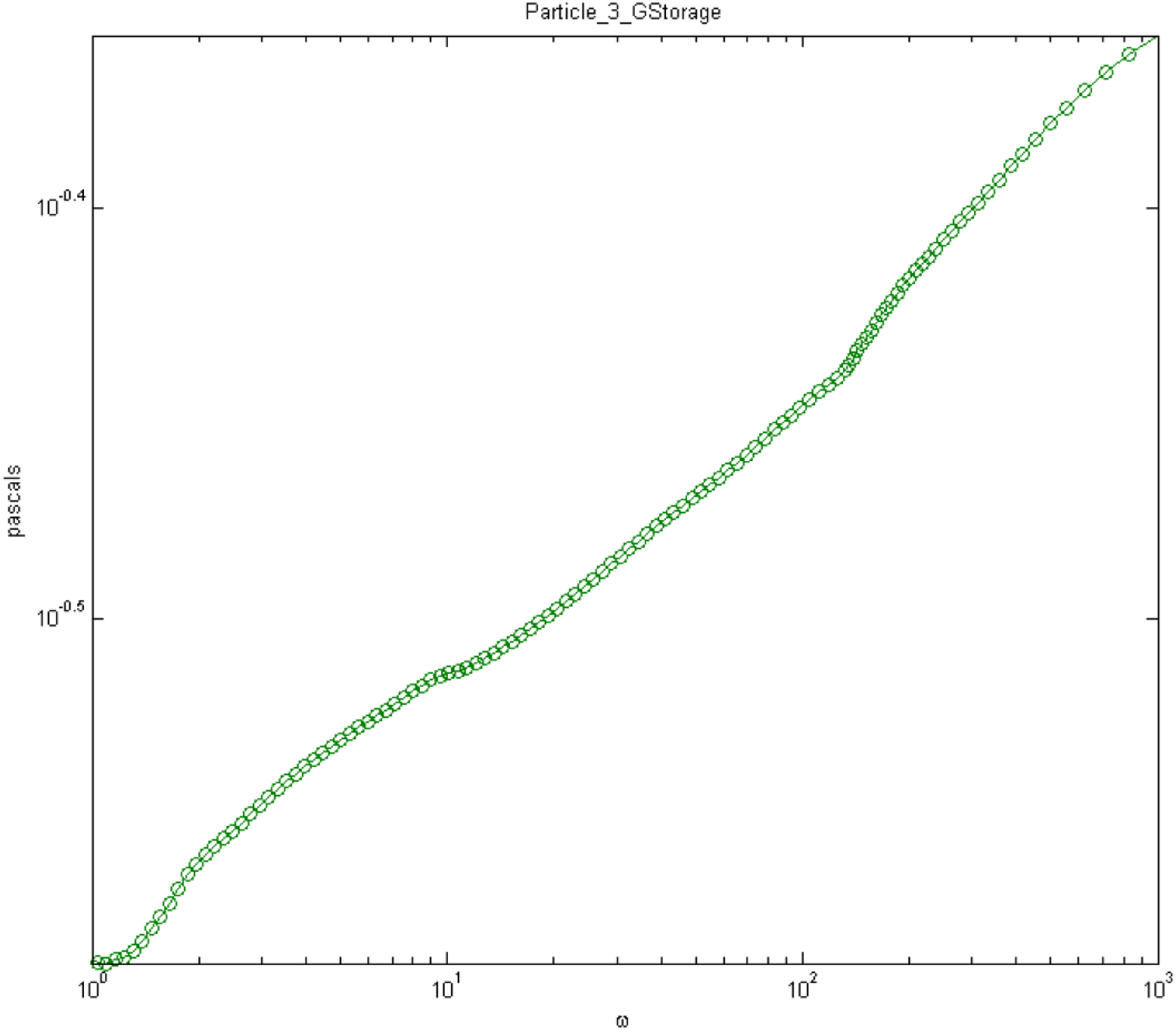}
\vfill

\Large{Exemplary Measurement: Distance to the rim of the nucleus 10.5~$\mu$m:}

\Large{Storage modulus}

\large{Supplemental data, Comparison of nanomechanical properties of in
vivo and in vitro keratin 8/18 networks Tobias Paust, Anke Leitner, Ulla Nolte, Michael Beil, Harald Herrmann, Othmar Marti}

\newpage

\includegraphics[width=\textwidth,height=0.8\textheight,keepaspectratio]{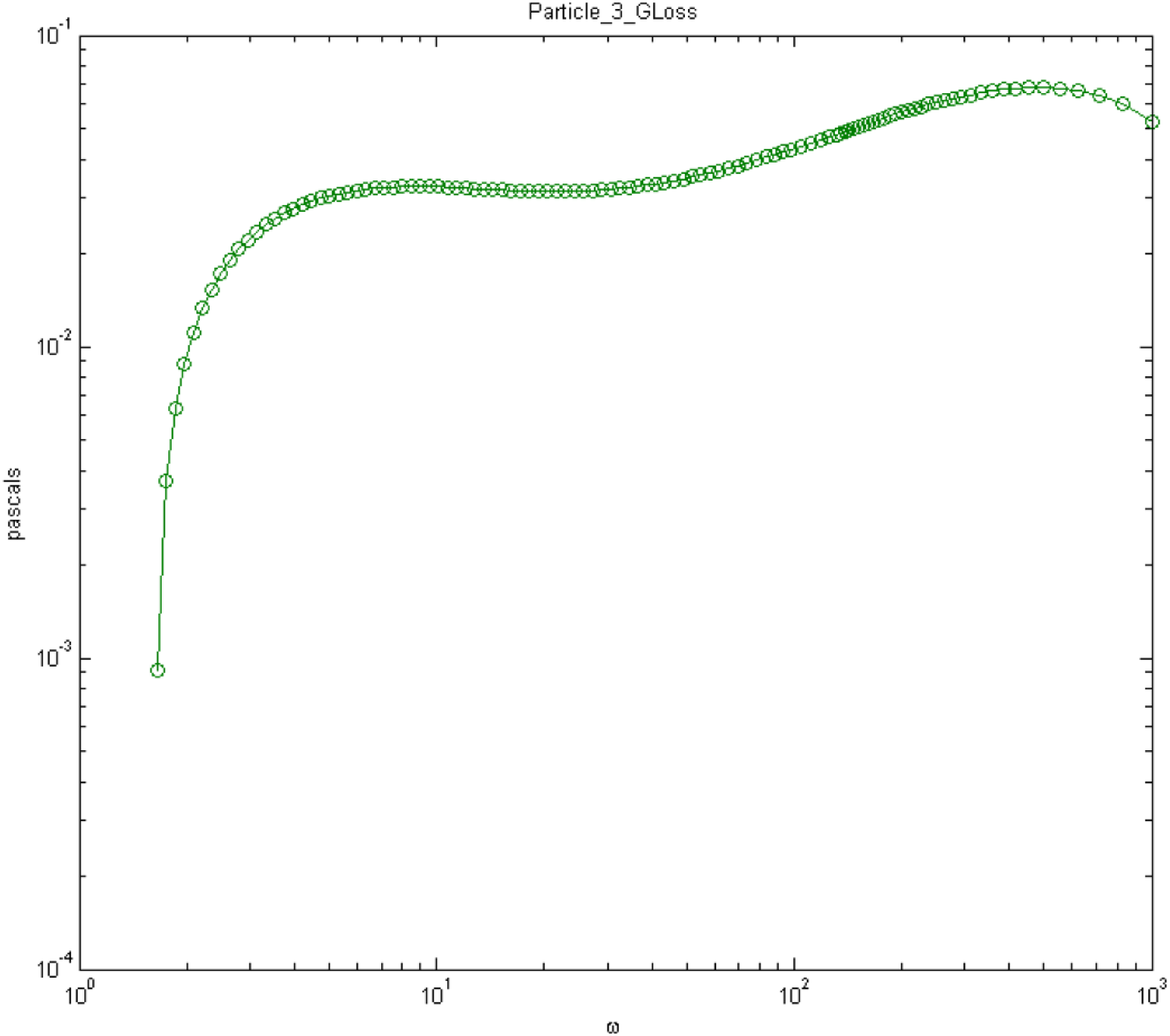}
\vfill

\Large{Exemplary Measurement: Distance to the rim of the nucleus 10.5~$\mu$m:}

\Large{Loss modulus}

\large{Supplemental data, Comparison of nanomechanical properties of in
vivo and in vitro keratin 8/18 networks Tobias Paust, Anke Leitner, Ulla Nolte, Michael Beil, Harald Herrmann, Othmar Marti}

\newpage

\includegraphics[width=\textwidth,height=0.8\textheight,keepaspectratio]{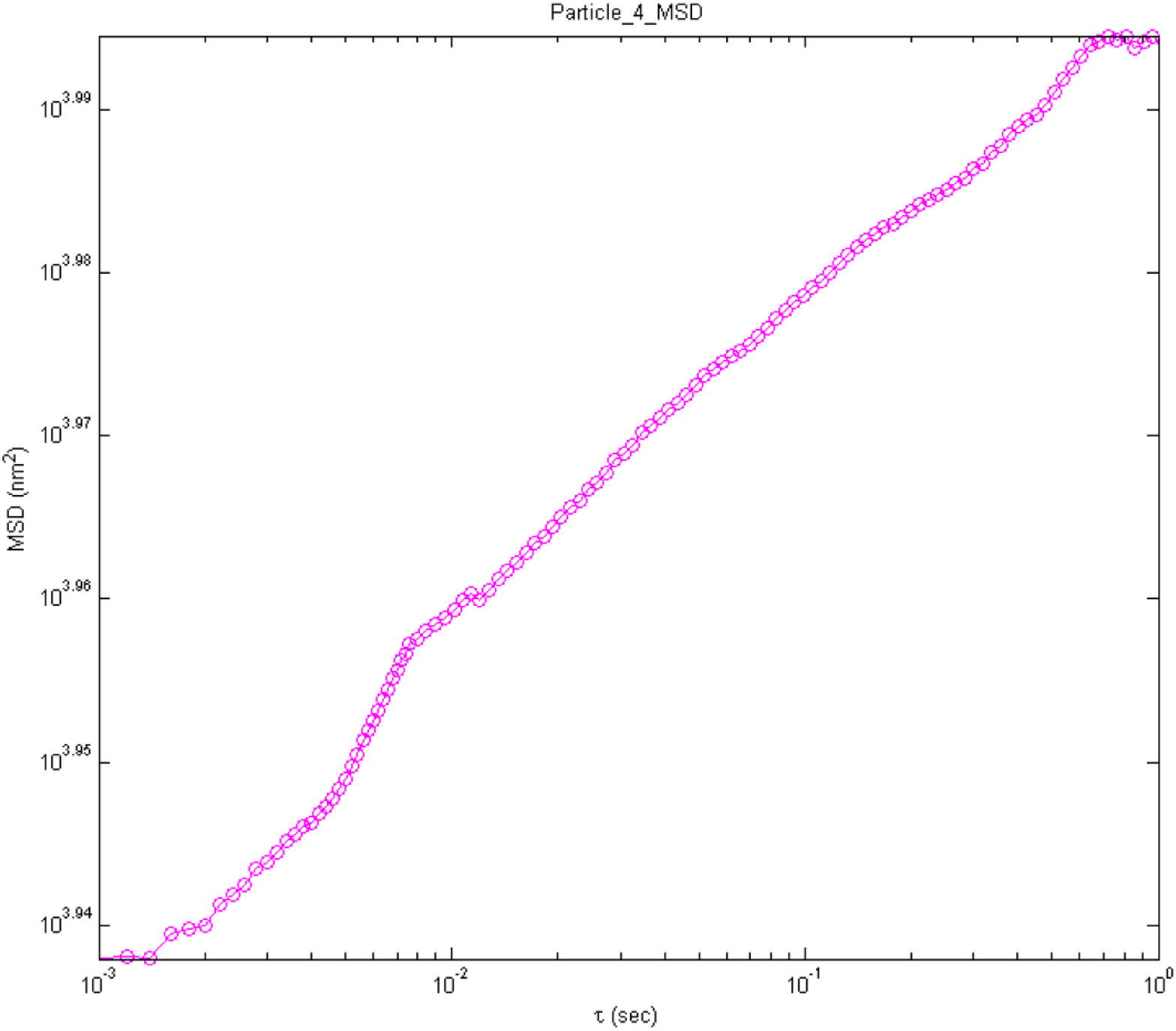}
\vfill

\Large{Exemplary Measurement: Distance to the rim of the nucleus 19.6~$\mu$m:}

\Large{Mean-square displacement}

\large{Supplemental data, Comparison of nanomechanical properties of in
vivo and in vitro keratin 8/18 networks Tobias Paust, Anke Leitner, Ulla Nolte, Michael Beil, Harald Herrmann, Othmar Marti}

\newpage

\includegraphics[width=\textwidth,height=0.8\textheight,keepaspectratio]{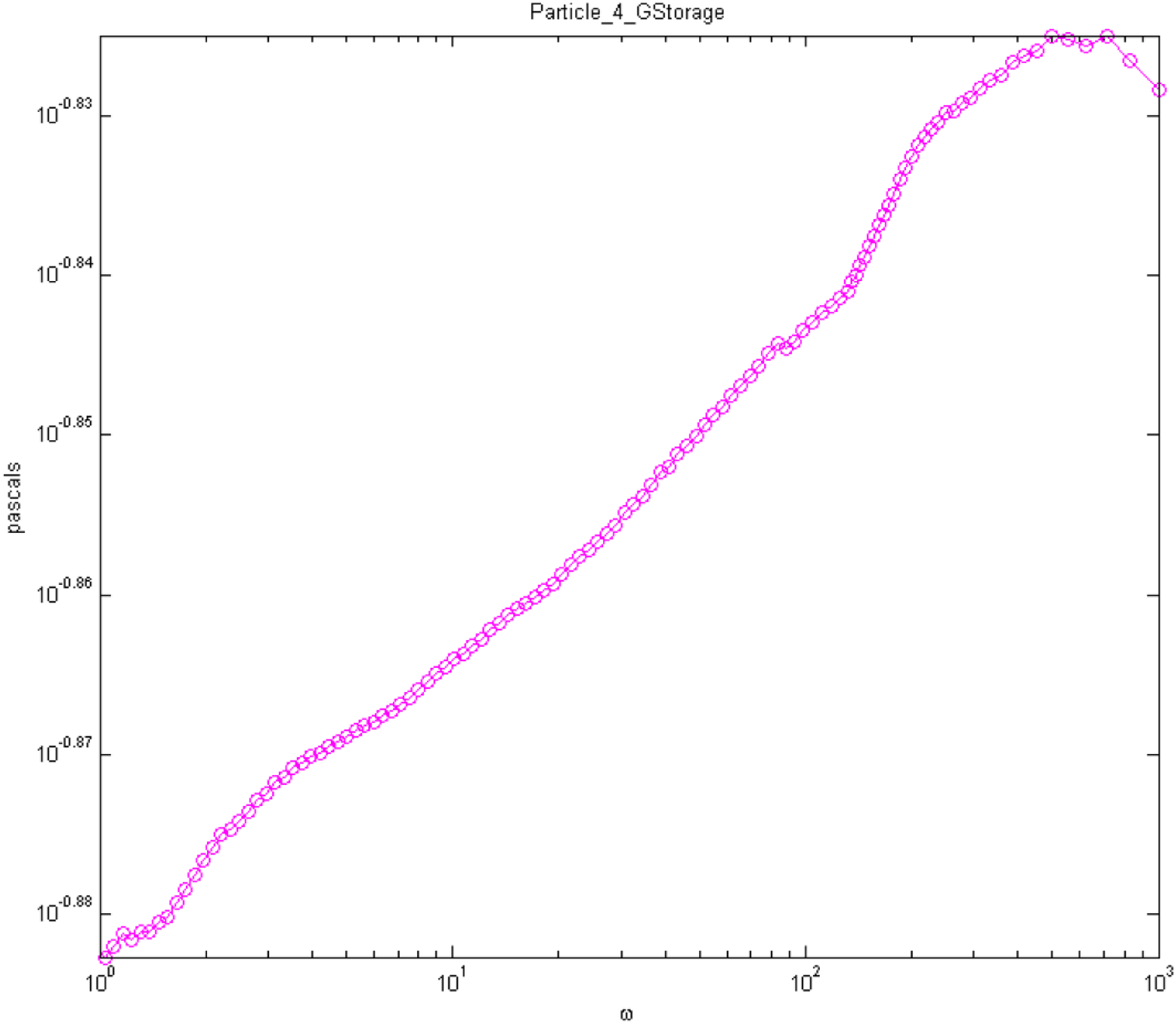}
\vfill

\Large{Exemplary Measurement: Distance to the rim of the nucleus 19.6~$\mu$m:}

\Large{Storage modulus}

\large{Supplemental data, Comparison of nanomechanical properties of in
vivo and in vitro keratin 8/18 networks Tobias Paust, Anke Leitner, Ulla Nolte, Michael Beil, Harald Herrmann, Othmar Marti}

\newpage

\includegraphics[width=\textwidth,height=0.8\textheight,keepaspectratio]{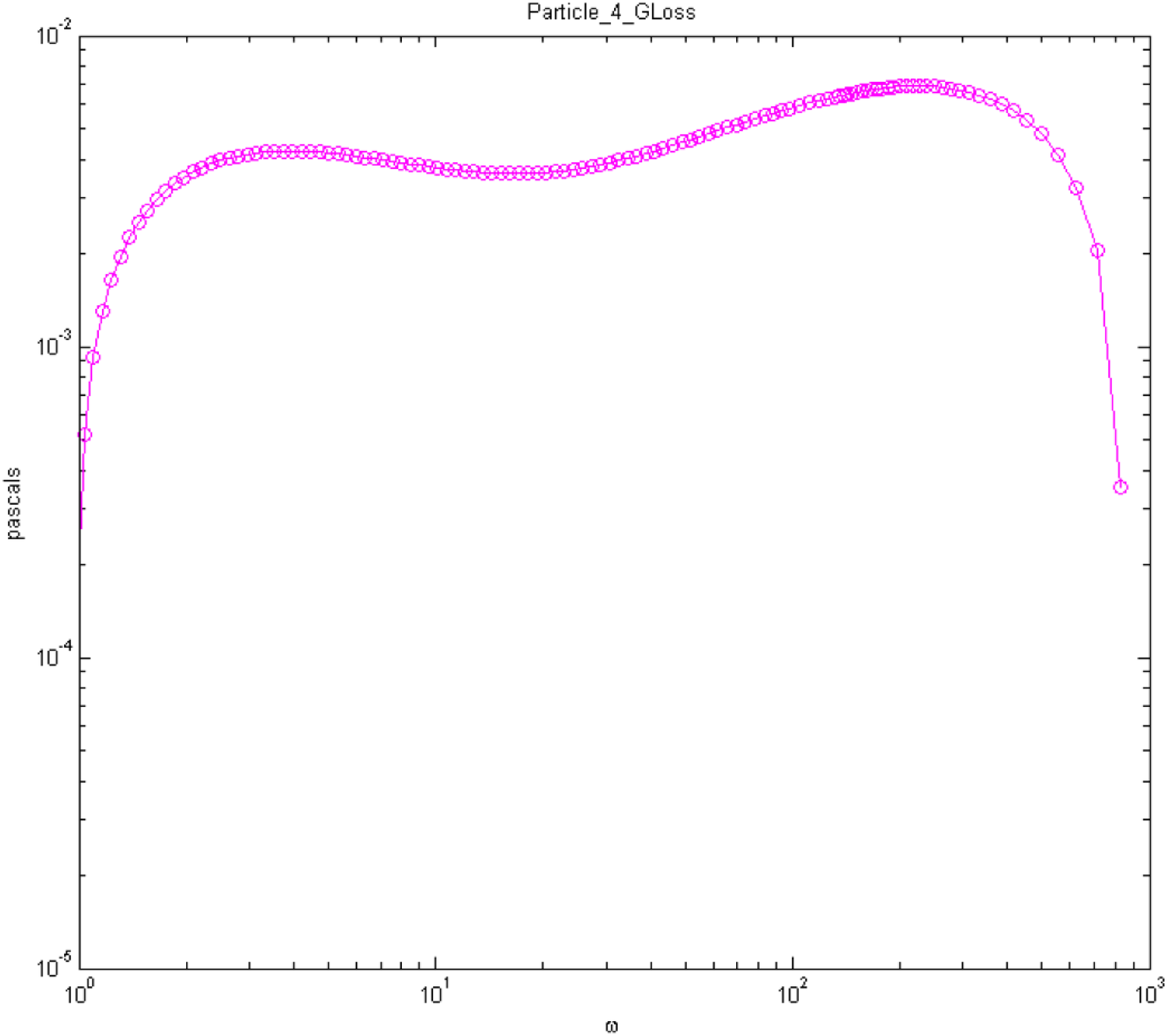}
\vfill

\Large{Exemplary Measurement: Distance to the rim of the nucleus 19.6~$\mu$m:}

\Large{Loss modulus}

\large{Supplemental data, Comparison of nanomechanical properties of in
vivo and in vitro keratin 8/18 networks Tobias Paust, Anke Leitner, Ulla Nolte, Michael Beil, Harald Herrmann, Othmar Marti}

\newpage

\includegraphics[width=\textwidth,height=0.8\textheight,keepaspectratio]{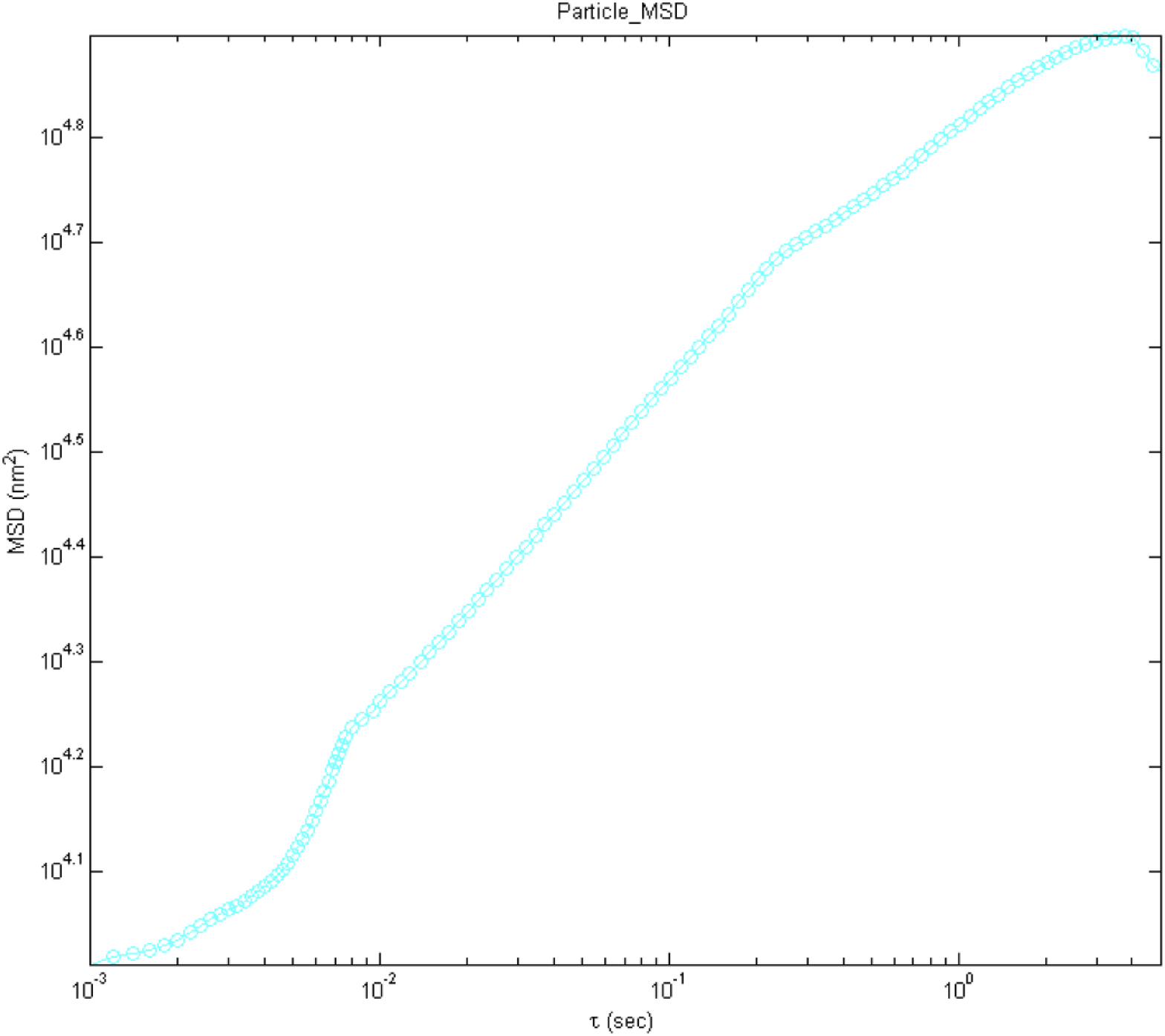}
\vfill

\Large{In vitro assembled keratin network: Example Measurement:}

\Large{Mean-square displacement}

\large{Supplemental data, Comparison of nanomechanical properties of in
vivo and in vitro keratin 8/18 networks Tobias Paust, Anke Leitner, Ulla Nolte, Michael Beil, Harald Herrmann, Othmar Marti}

\newpage

\includegraphics[width=\textwidth,height=0.8\textheight,keepaspectratio]{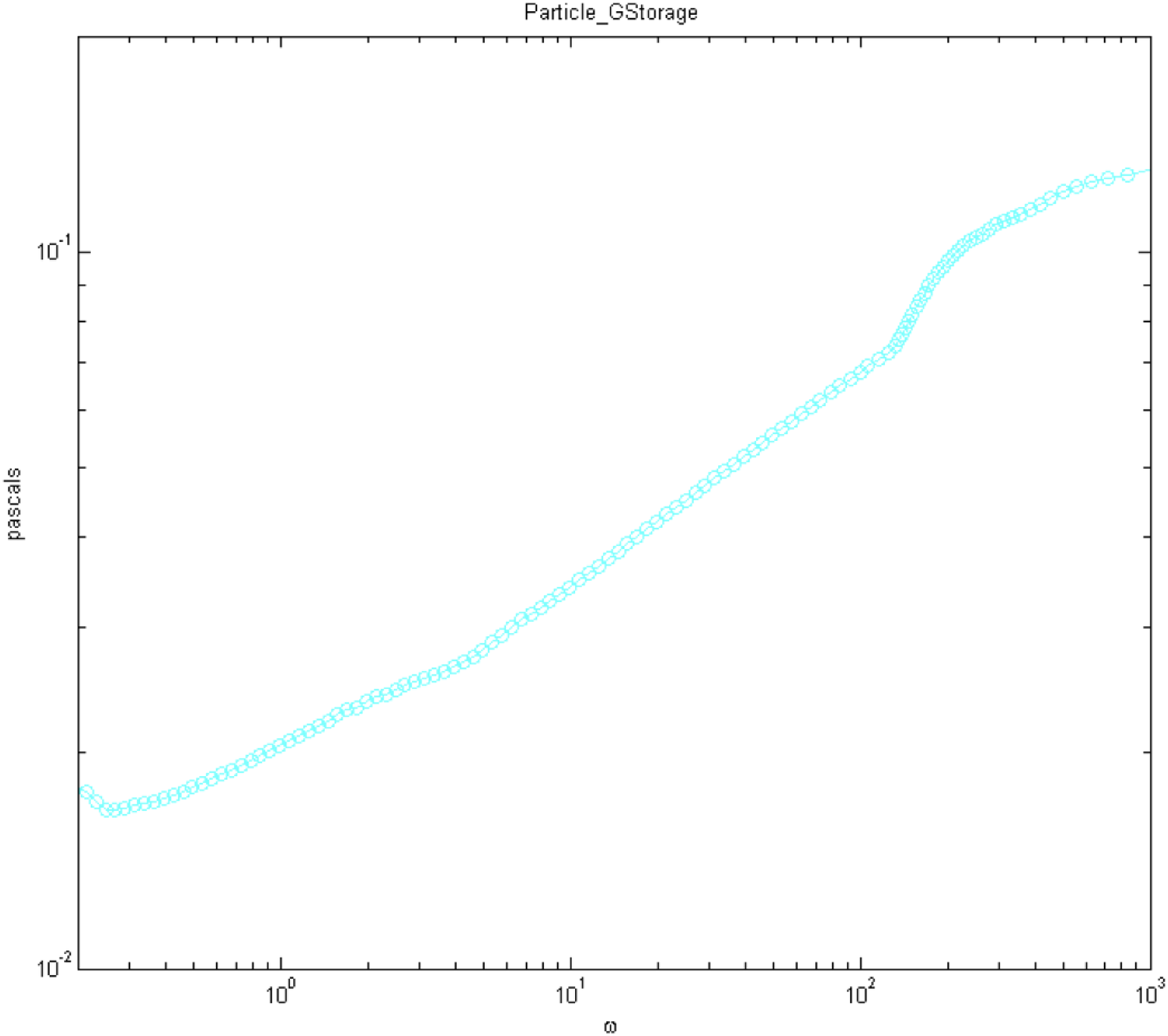}
\vfill

\Large{In vitro assembled keratin network: Example Measurement:}

\Large{Storage modulus}

\large{Supplemental data, Comparison of nanomechanical properties of in
vivo and in vitro keratin 8/18 networks Tobias Paust, Anke Leitner, Ulla Nolte, Michael Beil, Harald Herrmann, Othmar Marti}

\newpage

\includegraphics[width=\textwidth,height=0.8\textheight,keepaspectratio]{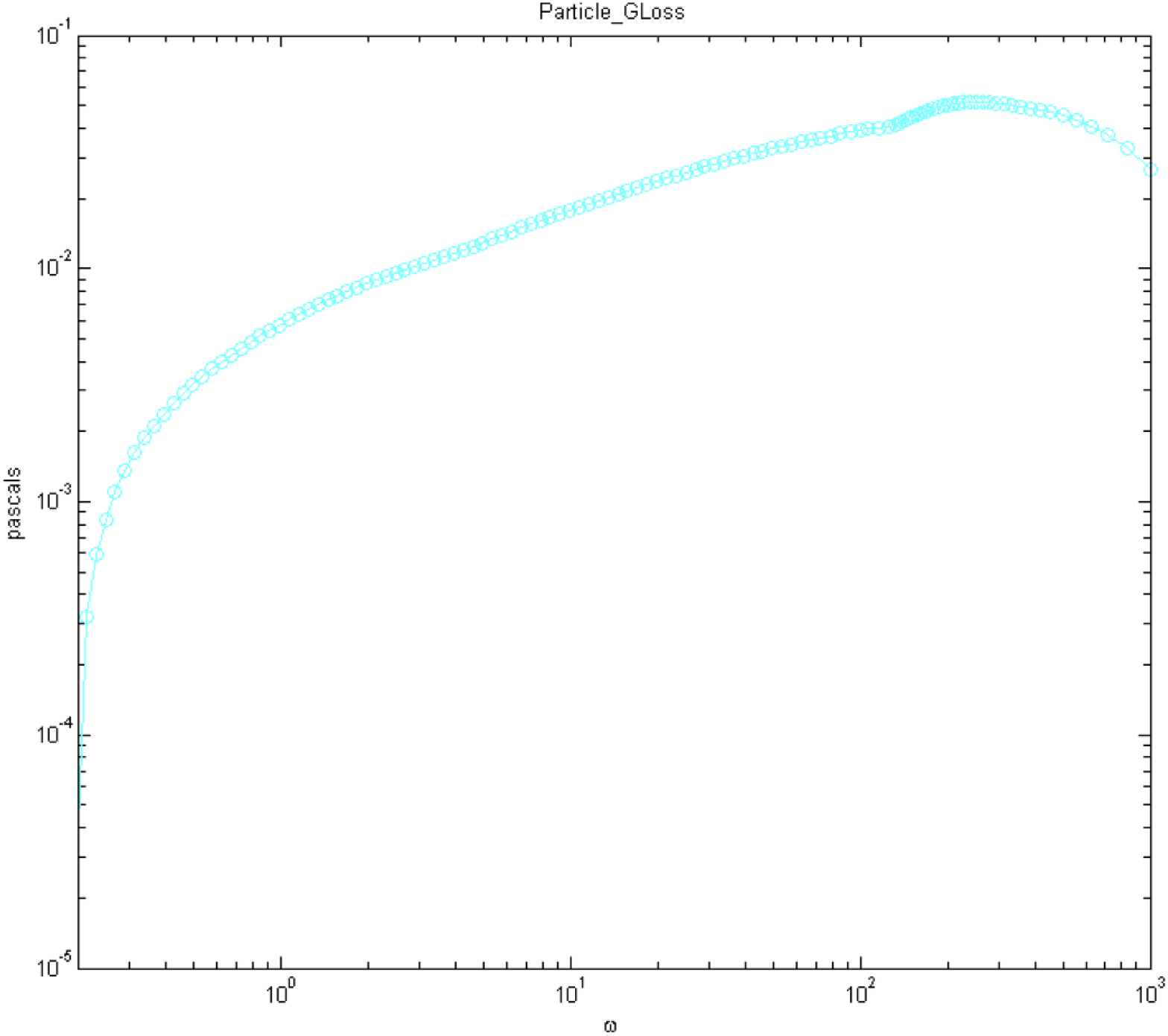}
\vfill

\Large{In vitro assembled keratin network: Example Measurement:}

\Large{Loss modulus}

\large{Supplemental data, Comparison of nanomechanical properties of in
vivo and in vitro keratin 8/18 networks Tobias Paust, Anke Leitner, Ulla Nolte, Michael Beil, Harald Herrmann, Othmar Marti}

\newpage

\includegraphics[width=\textwidth,height=0.8\textheight,keepaspectratio]{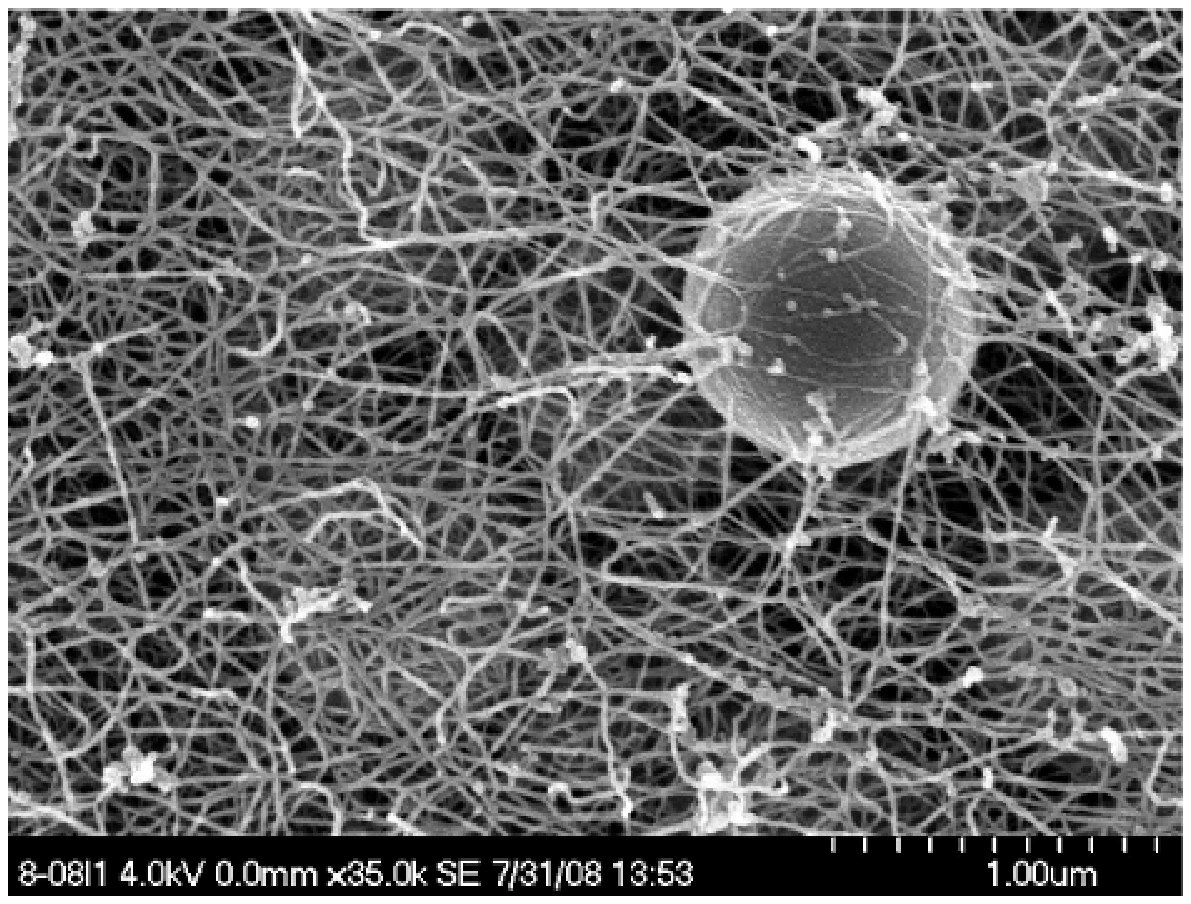}
\vfill

\Large{Electron microscopy: Picture of embedded bead in the extracted network}

\large{Supplemental data, Comparison of nanomechanical properties of in
vivo and in vitro keratin 8/18 networks Tobias Paust, Anke Leitner, Ulla Nolte, Michael Beil, Harald Herrmann, Othmar Marti}

\newpage

\includegraphics[width=\textwidth,height=0.8\textheight,keepaspectratio]{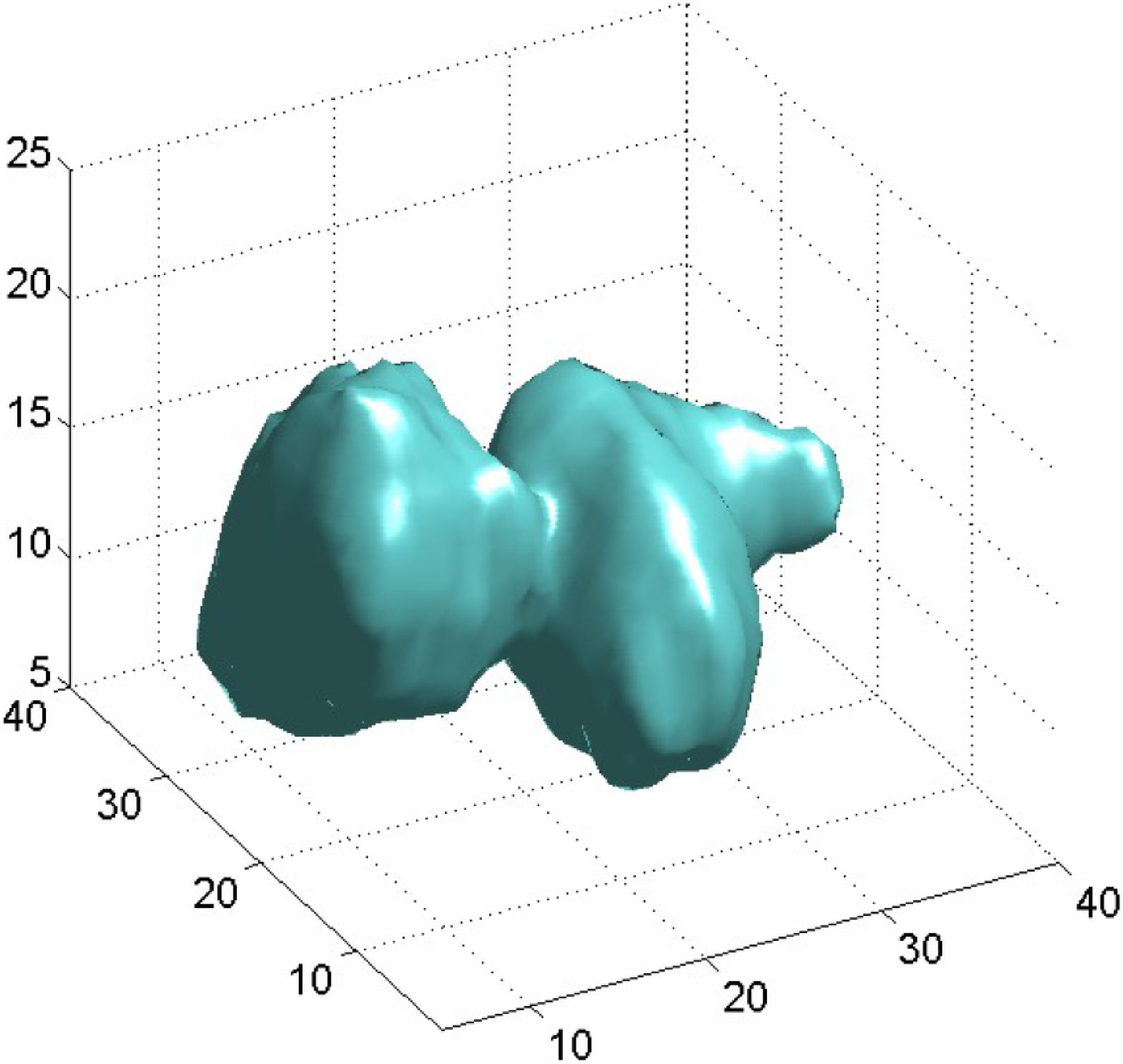}
\vfill

\Large{Iso surface representation of 3d position distribution in the in vitro assembled network:
1}

\large{Supplemental data, Comparison of nanomechanical properties of in
vivo and in vitro keratin 8/18 networks Tobias Paust, Anke Leitner, Ulla Nolte, Michael Beil, Harald Herrmann, Othmar Marti}

\end{document}